# Free Will and Advances in Cognitive Science

Leonid Perlovsky
Harvard University and AFRL, leonid@seas.harvard.edu

**Abstract –** Free will is fundamental to morality, intuition of self, and normal functioning of the society. However, science does not provide a clear logical foundation for this idea. This paper considers the fundamental scientific argument against free will, called reductionism, and explains the reasons for choosing dualism against monism. Then, the paper summarizes unexpected conclusions from recent discoveries in cognitive science. Classical logic turns out not to be the fundamental mechanism of mind. It is replaced by dynamic logic. Mathematical and experimental evidence are considered conceptually. Dynamic logic counters logical arguments for reductionism. Contemporary science of mind is not reducible; free will can be scientifically accepted along with scientific monism.

1. FREE WILL

The question of free will ranks amongst the three or four most important philosophical problems of all time (Catholic Encyclopedia). Yet, it cannot be reconciled with science. Most contemporary philosophers and scientists do not believe that free will exists (Bering 2010). Scientific arguments against the reality of free will can be summarized as follows (Wikipedia 2010a). Scientific method is fundamentally monistic: spiritual events, states, and processes (the mind) are to be explained based on laws of matter, from material states and processes in the brain. The basic premise of science is causality, future states are determined by current states, according to the laws of physics. If physical laws are deterministic then there is no free will, since determinism is opposite to freedom. If physical laws contain probabilistic elements or quantum indeterminacy, there is still no free will, since indeterminism and randomness are the opposites of freedom (Lim 2008; Bielfeldt 2009).

Free will, however, has a fundamental position in many cultures. Morality and judicial systems are based on free will. Denying free will threatens to destroy the entire social fabric of the society (Rychlak 1983; Glassman 1983). Free will also is a fundamental intuition of self. Most of the people on Earth would rather part with science than with the idea of free will (Bering 2010). Most people, including many philosophers and scientists, refuse to accept that their decisions are governed by the same laws of nature as a piece of rock by the road wayside or a leaf flown by the wind. (e.g. Libet 1999; Velmans 2003). Yet, the reconciliation of scientific causality and free will remains an unsolved problem.

This paper outlines a scientific theory reconciling free will and science. A scientific theory requires unambiguous predictions that can be experimentally verified. Such predictions and their experimental confirmations are discussed later.

2. MONISM AND DUALISM

The above arguments assume scientific monism, which states that the spiritual states of mind are produced by material processes in the brain. It seems that scientific monism, by accepting the unity of matter and spirit, fundamentally contradicts freedom of the will. This position of monism denying free will was accepted by B. Spinoza among many great thinkers (Wikipedia, 2010b). Other great thinkers could not accept this conclusion, rejected monism and chose dualism, according to which spiritual and material substances are different in principle and governed by different laws. Among famous dualists are R. Descartes

(Wikipedia, 2010c), and D. Chalmers (Wikipedia, 2010d). Under pressure from science and logic, many theologists reject the monotheistic interpretation of religions, despite this fundamental basis of Judaism, Buddhism, Christianity, and Islam.

Rejecting monism and accepting dualism (of matter and spirit) contradicts the fundamentals of our culture. Dualistic position attempts to separate the laws of spirit and the laws of matter, however there is no scientific principle to to accomplish this. Therefore, dualism cannot serve as a foundation of science. The only basis for separating the laws of spirit from the laws of matter, it seems, is to accept as material that which is currently explained by science, and declare as spiritual all that which seems unexplainable. Any hypothesis attempting such a separation of spirit and matter at any moment in history will be falsified by science many times over. The monistic view that spirit and matter are of the same substance is not only the basic foundation of science, but also corresponds to the fundamental theological positions of most world religions.

Logical constructions (e.g. dual-aspect frameworks) have been suggested that unify monism and dualism (Vimal 2009). These logical constructions however have not been able to point a direction for developing a scientific approach for such unification (that is, experimentally verifiable predictions, or how these could be approached; I would emphasize that this is the mainstay of monistic-materialistic science, such as physics).

The set of issues involving free will, monism and dualism, science, religions, and cultural traditions are difficult to reconcile (e.g. Chalmers 1995; Velmans 2008). The main difficulty is sometimes summarized as reductionism: if the highest spiritual values could be scientifically reduced to biological explanations, eventually they would be reduced to chemistry, to physics, and there would be no difference between laws governing the mind and spiritual values on the one hand, and a leaf flown with the wind on the other.

3. REDUCTIONISM AND LOGIC

Physical biology has explained the molecular foundations of life, DNA and proteins. Cognitive science has explained many *mental* processes in terms of *material* processes in the brain. Yet, molecular biology is far away from mathematical models relating processes in the mind to DNA and proteins. Cognitive science is only approaching some of the foundations of perception and simplest actions (Perlovsky 2006a). Nobody has ever been able to scientifically reduce the highest spiritual processes and values to the laws of physics. All reductionist arguments and difficulties of free will discussed above, when applied to highest spiritual processes, have not been based on mathematical predictive models with experimentally verifiable predictions—the essence and hallmark of science. All of these arguments and doubts were based on logical arguments. Logic has been considered a fundamental aspect of science since its very beginning and fundamental to human reason during more than two thousand years. Yet, no scientist will consider logical argument sufficient, in the absence of predictive scientific models, confirmed by experimental observations.

In the 1930s a mathematical logician Gödel (1934) discovered the fundamental deficiencies of logic. These deficiencies of logic are well known to scientists and are considered among the most fundamental mathematical results of the twentieth century. Nevertheless, logical arguments continue to exert powerful influence on scientists and non-scientists alike. Let me repeat the fact that most scientists to do not believe in free will. This rejection of fundamental cultural values and an intuition of self without scientific evidence seem to be a glaring contradiction. Of course, there have to be equally fundamental psychological reasons for such rejection, most likely originating in the subconscious. The rest of the paper analyzes these reasons and demonstrates that the discussed doubts are indeed unfounded. To understand the new arguments we will look into the recent evolution of cognitive science and mathematical models of the mind.

4. RECENT COGNITIVE THEORIES AND DATA

Attempts to develop mathematical models of the mind (computational intelligence) have for decades encountered irresolvable problems related to computational complexity. All developed approaches, including artificial intelligence, pattern recognition, neural networks, fuzzy logic and others faced complexity of computations, the number of operations exceeding the number of all elementary interactions in the universe (Bellman 1961; Minsky 1975; Winston 1984; Perlovsky 1998, 2001, 2006a,b). The mathematical analysis of this complexity problem related it to the difficulties of logic demonstrated by Gödel (1934). It turned out that complexity was a manifestation of Gödelian incompleteness in finite systems, such as computers or brains, (Perlovsky 1996, 2001). The difficulties of computational intelligence turned out to be related to the most fundamental mathematical result of the $20^{th}$ century.

A different type of logic was necessary for overcoming the difficulty of complexity. Dynamic logic is a process-logic, a process "from vague to crisp," from vague statements, conditions, models to crisp ones (Perlovsky, 1987, 1989, 2001, 2006a,b 2010b; Perlovsky & McManus1991). Dynamic logic is not a collection of static statements (such as 'this is a chair' or 'if A then B'); it is a dynamic logic-process. Dynamic logic was applied to solving a number of engineering problems that could not have been solved for decades because of the mathematical difficulties of complexity (Perlovsky 1989, 1994, 2001, 2004, 2007a,b,c,d, 2010b,d; Perlovsky, Chernick, & Schoendorf 1995; Perlovsky, Schoendorf, Burdick, & Tye 1997; Perlovsky et al 1997; Tikhanoff et al 2006; Perlovsky & Deming 2007; Deming & Perlovsky 2007; Perlovsky & Kozma 2007; Perlovsky & Mayorga 2008; Perlovsky & McManus 1991). These engineering breakthroughs became possible because dynamic logic mathematically models perception and cognition in the brain-mind. A basic property of dynamic logic is that it describes perception and cognition as processes in which vague (fuzzy) mental representations evolve into crisp representations. More generally, dynamic logic describes interaction between bottom-up and top-down signals (to simplify, signals from sensor organs, and signals from memory). Mental representations in memory, sources of top-down signals, are vague; during perception and cognition processes they interact with bottom-up signals, and evolve into crisp mental representations; crispness of the final states correspond to crispness of the bottom-up representations, e.g., retinal images of objects in front of our eyes. Initial vague representations and the dynamic logic process from vague to crisp are unconscious; only the final states, in which top-down representations match patterns in bottom-up signals, are available to consciousness and mentally perceived as approximately logical states.

During recent decades much became known about neural mechanisms of the mind-brain, especially about mechanisms of perception at the lower levels of the mental hierarchy (Grossberg 1988). This foundation makes it possible to verify the vagueness of initial states of mental representations. It is known that visual imagination is created by top-down signals. If one closes one's eyes, and imagines an object, seen just a second ago, this imagination gives an idea of properties of mental representations of the object. The imagined object is vague compared with the object perceived with opened eyes. If we open our eyes, it seems that we immediately perceive the object crisply and consciously. However, it is known that it takes approximately 160 ms to perceive the object crisply and consciously; therefore the neural mechanisms acting during these 160 ms are unconscious. This crude experimental verification of dynamic logic predictions was confirmed in detailed neuroimaging experiments (Bar et al 2006; Perlovsky 2009c). Mental representations in memory are vague and less conscious with closed eyes; with opened eyes they are not conscious. Opened eyes mask vagueness of initial mental states from our consciousness. Dynamic logic mathematically models a psychological theory of Perceptual Symbol System (Barsalou 1999; Perlovsky & Ilin 2010b). In this theory symbols in the brain are processes simulating experiences, and they are mathematically modeled by dynamic logic processes.

5. HIERARCHY OF THE MIND

The mind is organized into an approximate hierarchy. At the lower levels of the hierarchy we perceive sensory features. We do not normally experience free will with regard to functioning of our sensor systems. Higher up the mind perceives objects, still higher, situations, abstract concepts. Each next higher level contains more general and more abstract mental representations. These representations are built (learned) on top of lower level representations, and correspondingly, representations at every higher level are vaguer and less conscious (Perlovsky 2006a,c,d, 2007b,c; Perlovsky 2008; Perlovsky 2010a,c; Mayorga & Perlovsky 2008). For example, at a lower level the mind may perceive objects, such as a computer, a chair, a desk, bookshelves with books; each object is perceived due to a representation, which organizes perceptual features into the unified object; at this low level perception mechanisms function autonomously, mostly unconsciously, and free will is not experienced. At a higher level, the mind perceives a situation, say a professor's office, which is perceived due to a corresponding representation as an organized whole made up of objects. We experience free will about, say moving objects and arranging furniture in our office. At still higher levels the mind cognizes ideas of a University, or a system of education, due to representations at corresponding levels. And even if we appreciate that an individual ability of changing educational system might be limited, still we experience free will to think about it. Whereas in everyday mundane experience we know that our freedom is limited in many ways, still, at higher levels of the mind we experience intuitions or ideas of free will and self, possessing free will.

Many people doubt that free will exists, for the reasons of scientific causality and reducibility discussed above. Therefore I remind that even at the level of simple object perception, mental representations (in absence of actual objects) are vague and barely conscious. Higher up, on top of several vague and less conscious levels of the hierarchy, contents of representations are vague and unconscious. However, believing in free will, despite severe limitations of our freedom in real life, consciously or unconsciously, is extremely important for individual survival, for achieving higher goals, and for evolution of cultures (Glassman 1987; Bielfeldt 2009). In animal kingdom "belief in free will" acts instinctively, their psyche is unified. Similarly this question did not appear in the mind of our early progenitors. In human mind, for hundreds of thousands of years belief in free will directed actions of early humans unconsciously. An intuition of free will is a recent cultural achievement. For example, in Homer Iliad, only Gods possess free will; 100 years later Ulysses demonstrates a lot of free will (Jaynes 1976). Clearly, conscious contemplation of free will is a cultural construct. It became necessary with evolution and the differentiation of consciousness and culture. The majority of cultures existing today have well developed ideas about free will, religious and educational systems for installing these ideas in the minds of every next generation. But does free will really exist? To answer this question, and even to understand the meaning of *really* we will now consider how ideas exist in culture, and how the existence of ideas in cultural consciousness differs from ideas in individual cognition (cultural consciousness refers to what is conscious in culture, in its texts, practices, etc.).

6. LANGUAGE AND COGNITION

Cultures accumulate knowledge and transmit it from generation to generation mostly due to language. Mechanisms of interactions between language and cognition (Perlovsky 2004; 2007e; 2009a,b; Fontanari & Perlovsky 2007, 2008a,b; Perlovsky & Ilin 2010a,b) explain why language is acquired in childhood, whereas higher cognition requires much longer acquisition time. How do we learn the correct connections between words and objects, among the multitude of incorrect ones (no amount of experience would be sufficient to overcome computational complexity of learning these connections)? Why does not human-level cognition evolve in animals without language? What, exactly, are the similarities and differences between language and cognition?

Following the given references, these and other properties of cognition-language interaction are explained according to the mechanism of the dual model hierarchy (Fig. 1). This figure illustrates the dual hierarchy of the mind, a cognitive hierarchy from sensory signals, to objects, to situations, to abstract concepts… and a parallel hierarchy of language from words, to phrases, from concrete to abstract meanings.

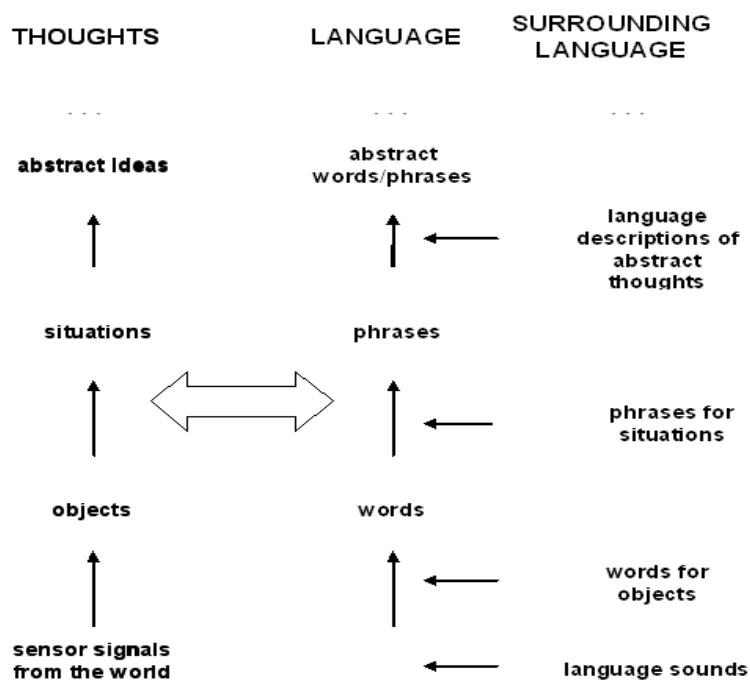

Fig.1. Parallel hierarchies of cognition and language. Language learning is grounded in surrounding language at all hierarchical levels ready-made. Learning abstract cognitive models requires experience and guidance from language.

The dual model along with dynamic logic suggests that a newborn brain contains separate place-holders for future representations of language and cognitive contents. The initial contents are vague and non-specific. The newborn mind has no image-representations, say for chairs, or sound-representations for an English word *chair*. Yet, the connections between placeholders for future cognitive and language representations are inborn. The inborn connections between cognitive and language brain areas are not surprising, Arbib (2005) suggested that such connections existed due to the mechanism of mirror neurons millions of years before language ability evolved. Due to these inborn connections, word and object representations are acquired correctly connected: as one part of the dual model (a word or object representation) is learned, becomes crisper and more specific, the other part of the dual model is learned in correspondence with the first one. Objects that are directly observed can be learned without language (like in animals). However, abstract ideas cannot be directly observed; they cannot be learned from experience as useful combinations of objects, because of computational complexity of such learning. Therefore, cognitive representations of abstract ideas can be learned from experience only due to guidance by language.

Language can be learned from surrounding language without real-life experience, because it exists in the surrounding language ready-made at all levels of the mind hierarchy. This is the reason language is acquired in childhood, whereas learning corresponding cognitive representations requires much more experience. Learning language can proceed fast, because it is grounded in surrounding language at all hierarchical levels. But cognition is grounded in direct experience only at the bottom levels of perception. At higher levels of abstract ideas, learning cognitive representations from experience is guided by already learned language representations. Abstract ideas that do not exist in language (in culture or in personal language) usually cannot be perceived or cognized and their existence are not noticed, until they are learned in language.

Language grounds and supports learning of the corresponding cognitive representations, similar to the eye supporting learning of an object representation in the opened-closed eye experiment. Language serves as inner mental eyes for abstract ideas. The fundamental difference, however, is that language 'eyes' cannot be closed at will. The crisp and conscious language eyes mask vague and barely conscious cognitive representations. Therefore we cannot perceive them. If we do not have the necessary experience, our cognitive representations are vague and unconscious and language representations are taken for this abstract knowledge. It is obvious with children, but it also persists through life. Because language contains wealth of cultural information, we are capable of reasonable judgments, even without direct life experience.

This discussion is directly relevant to Maimonides' interpretation of the Original Sin (Levine & Perlovsky 2008), Adam was expelled from paradise because he did not want to think, but ate from the tree of knowledge to acquire existing knowledge ready-made. In terms of Fig.1, he acquired language knowledge from surrounding language but not in cognitive representations from his own experience. This discussion is also directly relevant to the difference between much discussed (Noble Prize 2002) irrational heuristic decision-making discovered by Tversky & Kahneman (1974, 1981) and decision-making based on personal experience and careful thinking, grounded in learning and driven by the knowledge instinct (Levine & Perlovsky, 2008; Perlovsky, Bonniot-Cabanac, & Cabanac 2010). In those cases when life experience is insufficient and cognitive representations are vague, crisp and conscious language representations substitute for the cognitive ones. This substitution is smooth and unconscious, so that we do not notice (without specific scientific training) when we speak from real life experience, or from language-based knowledge (heuristics). Language-based knowledge accumulates millennial wisdom and could be very good, but it is not the same as personal cognitive knowledge combining cultural wisdom with life experience. It might sound tautologically that we are conscious only about consciousness, and unconscious about unconsciousness. But it is not a tautology that we have no idea of nearly 99% of our mind functioning. Our consciousness jumps from one tiny conscious and logical island in our mind to another one, across an ocean of vague unconscious, yet our consciousness keeps 'us' sure that we are conscious all the time, and that logic is a fundamental mechanism of perception and cognition. Because of this property of consciousness, even after Gödel, most scientists have remained sure that logic is the main mechanism of the mind.

Return now to the question, does free will really exist? The question whether free will exists in the sense of resolving the free-will vs. determinism debate exists in classical logic, but it does not exist as a fundamental scientific question. Because of the properties of mental representations near the top of the mind hierarchy this question cannot be formulated within classical logic.

How can the question about free will be answered within the developed theory of the mind? Free will does not exist in inanimate matter. Free will exists as a cultural concept. The contents of this concept include all related discussions in cultural texts, literature, poetry, art, in cultural norms. This cultural knowledge gives the basis for developing corresponding language representations in individual minds; language representations are mostly conscious. Clearly, individuals differ by how much cultural contents they acquire from surrounding language and culture. The dual model suggests that, based on this personal language representation of free will, every individual develops his or her personal cognitive representation of this idea, which assembles his or her related experiences in real life, language, thinking, acting, into a coherent whole.

7. CONCLUSION

The contents of cognitive representation of free will determine personal thinking, responsibility, will, and actions, which one exercises in his or her life. Clearly, due to a hierarchy of vague representations, the concept of free will is far removed from physical laws controlling molecular interactions. Therefore, logical arguments about reducibility are plainly wrong. Logic is not a fundamental mechanism of the mind. Mathematical details of the corresponding cognitive models, supporting experimental evidence, and future directions of experimental and theoretical research are discussed in the given references. Among these directions for future research are experimental verification of interaction between language and cognition. Psychological and neuroimaging experiments shall be used to confirm that language and cognitive representations are neurally connected before either of them becomes crisp; high level abstract ideas first become conscious and crisp in language, and then gradually become conscious and crisp in cognition; language representations are crisp and conscious long before cognitive representations become equally crisp and conscious; the higher up in the mental hierarchy the vaguer and less conscious are cognitive representations; many abstract cognitive representations remain vague and unconscious throughout life, even though people can fluently talk about them. Some of these ideas are being experimentally tested, and have received partial support.

This paper addressed a fundamental philosophical issue of how one could scientifically accept an idea of free will, while humans are collections of atoms and molecules having no freedom. For centuries this consideration has been propelled by logic toward the idea of reductionism, logically denying a possibility of free will. We explained belief in logic in many scientists and philosophers, even in those well familiar with Gödelian theory, by fundamental properties of consciousness: we are conscious only about logical or near logical states of the mind. We resolved this difficulty by pointing out that logic, although prominent in consciousness, is not a fundamental mechanism of the mind. Dynamic logic, proven experimentally, models the human mind as an approximate hierarchy of vaguer and vaguer representations. This model eliminates logical arguments of reductionism (supporting those scientists denying it earlier) and supports the agreement between free will, scientific monism, and science.